\documentclass[aip,apl,preprint,english,showkeys,showpacs,preprintnumbers,amsmath,amssymb,floatfix,superscriptaddress]{revtex4-1}
\usepackage[T1]{fontenc}
\usepackage[latin9]{inputenc}
\usepackage{textcomp}
\usepackage{amsbsy}
\usepackage{graphicx}
\usepackage{amssymb}
\usepackage{amsmath}
\usepackage{esint}
\makeatletter

\providecommand{\LyX}{L\kern-.1667em\lower.25em\hbox{Y}\kern-.125emX\@}

\@ifundefined{textcolor}{}
{%
 \definecolor{BLACK}{gray}{0}
 \definecolor{WHITE}{gray}{1}
 \definecolor{RED}{rgb}{1,0,0}
 \definecolor{GREEN}{rgb}{0,1,0}
 \definecolor{BLUE}{rgb}{0,0,1}
 \definecolor{CYAN}{cmyk}{1,0,0,0}
 \definecolor{MAGENTA}{cmyk}{0,1,0,0}
 \definecolor{YELLOW}{cmyk}{0,0,1,0}
 }

\makeatother

\usepackage{babel}

\begin{document}

\title{Lock-in detection for pulsed electrically detected magnetic resonance}

\author{Felix Hoehne}
\email[corresponding author, email: ]{hoehne@wsi.tum.de} \affiliation{Walter
Schottky Institut, Technische Universit\"{a}t M\"{u}nchen, Am Coulombwall 4,
85748 Garching, Germany}

\author{Lukas Dreher}
\affiliation{Walter Schottky Institut, Technische Universit\"{a}t
M\"{u}nchen, Am Coulombwall 4, 85748 Garching, Germany}

\author{Jan Behrends}
\altaffiliation{present address: Fachbereich Physik, Freie Universit\"at Berlin, Arnimallee 14, 14195 Berlin, Germany}
\affiliation{Helmholtz-Zentrum Berlin f\"ur Materialien und Energie, Institut f\"ur Silizium-Photovoltaik, Kekul\'estr. 5, 12489 Berlin, Germany}

\author{Matthias Fehr}
\email[email: ]{matthias.fehr@helmholtz-berlin.de}
\affiliation{Helmholtz-Zentrum Berlin f\"ur Materialien und Energie, Institut f\"ur Silizium-Photovoltaik, Kekul\'estr. 5, 12489 Berlin, Germany}

\author{Hans Huebl}
\affiliation{Walther-Mei\ss ner-Institut, Bayerische Akademie der Wissenschaften, Walther-Mei\ss ner-Str.\,8, 85748 Garching, Germany}

\author{Klaus Lips}
\affiliation{Helmholtz-Zentrum Berlin f\"ur Materialien und Energie, Institut f\"ur Silizium-Photovoltaik, Kekul\'estr. 5, 12489 Berlin, Germany}

\author{Alexander Schnegg}
\affiliation{Helmholtz-Zentrum Berlin f\"ur Materialien und Energie, Institut f\"ur Silizium-Photovoltaik, Kekul\'estr. 5, 12489 Berlin, Germany}

\author{Max Suckert}
\affiliation{Walter Schottky Institut, Technische Universit\"{a}t
M\"{u}nchen, Am Coulombwall 4, 85748 Garching, Germany}

\author{Martin Stutzmann}
\affiliation{Walter Schottky Institut, Technische Universit\"{a}t
M\"{u}nchen, Am Coulombwall 4, 85748 Garching, Germany}

\author{Martin S.~Brandt}
\affiliation{Walter Schottky Institut, Technische Universit\"{a}t
M\"{u}nchen, Am Coulombwall 4, 85748 Garching, Germany}
\begin{abstract}
We show that in pulsed electrically detected magnetic resonance (pEDMR) signal modulation in combination with a lock-in detection scheme can reduce the low-frequency noise level by one order of magnitude and in addition removes the microwave-induced non-resonant background. This is exemplarily demonstrated for spin-echo measurements in phosphorus-doped Silicon. The modulation of the signal is achieved by cycling the phase of the projection pulse used in pEDMR for the read-out of the spin state.  
\end{abstract}
\maketitle
Electron paramagnetic resonance (EPR) has proven to be a powerful tool in the characterization of defects in semiconductors~\cite{Spaeth03}. However, EPR is rather insensitive and typically limited to samples with more than 10$^{10}$ spins~\cite{Maier97}.
Due to its higher sensitivity, electrically detected magnetic resonance (EDMR) is now widely used to study defects, in particular in indirect, disordered or organic semiconductors~\cite{Lepine72Spindep, Frankevich84, Dyakonov1994, Carlos1997, Stutzmann2000, Graeff2005}. Over the last years, pulsed EDMR (pEDMR) has gained considerable interest, since it combines the large toolbox of pulsed EPR methods~\cite{Schweiger01} with the enhanced sensitivity of EDMR e.g.~to identify spin-dependent transport and recombination processes and study hyperfine interactions~\cite{Boehme03EDMR, Stegner06, Harneit2007, McCamey08, Behrends09II, Behrends10, HoehneENDOR2011, HoehneESEEM2011, Fehr2011}. However, in many cases pEDMR suffers from strong low-frequency noise and large non-resonant background signals induced by the strong microwave pulses used to manipulate the spin system~\cite{Stegner06}. Here, we demonstrate that for pEDMR, a lock-in detection scheme is able to subtract the non-resonant background and effectively reduce low-frequency noise by more than one order of magnitude following similar ideas that have been applied in conventional pulsed EPR spectroscopy~\cite{Percival75}.    

\begin{figure}[ht]
\begin{centering}
\includegraphics[width=8cm]{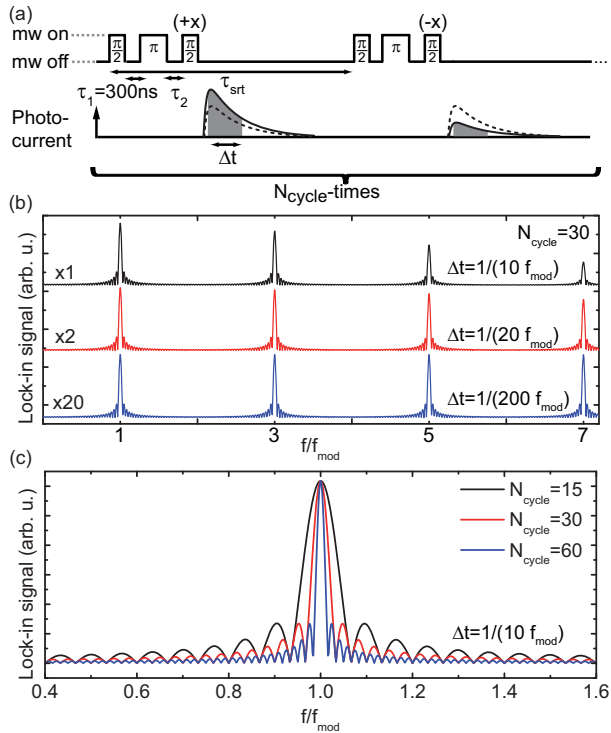}
\par\end{centering}
\caption{\label{fig:figure1}
(a) Pulse sequence to measure electrically detected spin echoes. For signal modulation, we alternately apply the spin-echo pulse-sequence with the phase of the last $\pi$/2 pulse set to (+x) and with its phase set to (\hbox{-}x). This cycle is repeated $N_\mathrm{cycle}$-times. The current transients (solid line) after the mw pulses consist of a spin-independent non-resonant (dashed line) and a spin-dependent resonant part. After the (\hbox{-}x) spin-echo pulse-sequence the resonant contribution to the current transient is inverted when compared to the current transient after the (+x) pulse sequence. The shaded area indicates the box-car integration interval $\Delta t$.   
(b) Calculated response of the lock-in detection scheme $\bar{h}(f)$ for different box-car integration intervals $\Delta t$ scaled by the indicated factors. 
(c) Bandwidth calculated for different numbers of cycles $N_\mathrm{cycle}$.}
\end{figure}

In the pulsed EDMR discussed here, the symmetry of a spin pair is changed by resonant microwave pulses resulting in a change of the recombination rate of excess carriers, which is reflected in a current transient after the microwave pulses. 
The pEDMR signal is obtained by box-car integrating the current transient after the pulse sequence over a time interval $\Delta t$, resulting in a charge $\Delta Q$ proportional to the recombination rate at the end of the pulse sequence~\cite{Boehme03EDMR}, as schematically shown in Fig.~\ref{fig:figure1}~(a). However, the strong microwave pulses also cause spin-independent non-resonant changes of the current due to e.g. rectification in the semiconductor sample resulting in additional noise and background signals which are typically much larger than the spin-dependent signals.
These effects can be mitigated by using a lock-in detection scheme for pEDMR measurements, as will be described in the following.

Lock-in detection employs modulation of a signal at a certain frequency and its phase-sensitive detection in combination with bandpass filtering~\cite{Dicke1946}.
We will discuss how such a scheme can be implemented in pEDMR exemplarily for the measurement of electrically detected spin echoes. We use a $\pi$/2-$\tau_1$-$\pi$-$\tau_2$-$\pi$/2 spin-echo pulse-sequence, where $\pi/2$ and $\pi$ denote microwave pulses with corresponding flipping angles and $\tau_1$ and $\tau_2$ denote the duration of periods of free evolution [Fig.~\ref{fig:figure1}~(a)]\,\cite{Huebl08Echo}.
Depending on the phase of the projection pulse (indicated in Fig.~\ref{fig:figure1} by $\pm$x), the detection echo-sequence forms an effective
2$\pi$ pulse for (+x) or an effective $\pi$ pulse for (\hbox{-}x), since a phase change of 180$^\circ$ results in a
reversed sense of rotation of the spin states on the Bloch sphere. Thus, the echo amplitude for a (\hbox{-}x) projection pulse is inverted when compared to a (+x) projection pulse. By repeating the spin echo pulse sequence $N_\mathrm{cycle}$ times with a shot repetition time $\tau_\mathrm{srt}$ and inverting the phase for every shot, the signal is square-wave modulated at a frequency $f_\mathrm{mod}=1/(2\tau_\mathrm{srt}$). For phase-sensitive detection, the $\Delta Q$ detected for (+x) and (\hbox{-}x) are multiplied by +1 and -1, respectively, and the result is averaged over all cycles. As shown below, this scheme is only sensitive to signals within a bandwidth $\Delta f=1/(2N_\mathrm{cycle}\tau_\mathrm{srt})=1/T_\mathrm{meas}$ around the modulation frequency $f_\mathrm{mod}$, where $T_\mathrm{meas}$ denotes the overall measurement time.

In contrast to conventional lock-in detection schemes, the signal in pEDMR is integrated only over the time interval $\Delta t$ which is typically much smaller than the shot repetition time $\tau_\mathrm{srt}=1/(2f_\mathrm{mod}$). We therefore calculate 
the response $h(f)$ of the detection scheme including the box-car integration interval $\Delta t$ for an input signal of the form $\sin(2\pi f t+\phi)$ representing a noise component with frequency $f$ and random phase $\phi$. The function $h(f,\phi)$ is given by  
\begin{equation}
\begin{split}
h(f,\phi) = & \frac{1}{N_\mathrm{cycle}} \sum^{N_\mathrm{cycle}-1}_{n=0} \left[\int^{2n \tau_\mathrm{srt}+\Delta t}_{2n \tau_\mathrm{srt}}\sin(2\pi f t+\phi) \mathrm{d}t - \right.\\
 & \left. \int^{(2n+1) \tau_\mathrm{srt}+\Delta t}_{(2n+1) \tau_\mathrm{srt}}\sin(2\pi f t+\phi) \mathrm{d}t\right].
 \end{split}  
\end{equation}
Since the phase of the noise signal is random, the response $h(f,\phi)$ has to be averaged over $\phi$, giving 
\begin{eqnarray}
\nonumber
 \bar{h}(f) & = & \sqrt{\frac{1}{2\pi}\int^{2\pi}_{0}h(f,\phi)^2 \mathrm{d}\phi} \\
  & = & \left|{\frac{\sin(\pi f \Delta t) \sin(2\pi f N_\mathrm{cycle} \tau_\mathrm{srt})}{\sqrt{2}\pi f N_\mathrm{cycle} \cos(\pi f \tau_\mathrm{srt})}}\right| .
\label{eq:2}
\end{eqnarray} 
The function $\bar{h}(f)$ is plotted in Fig.~\ref{fig:figure1} (b) for different box-car integration intervals $\Delta t$=1/(10$f_\mathrm{mod}$), 1/(20$f_\mathrm{mod}$) and 1/(200$f_\mathrm{mod}$) with $N_\mathrm{cycle}=30$. The lock-in detection scheme is only sensitive to signals at odd harmonics of $f_\mathrm{mod}$. For longer integration intervals $\Delta t$, the higher harmonics are suppressed when compared to the fundamental frequency while for short $\Delta t$ supression is not effective as can be seen for $\Delta t$=1/(200$f_\mathrm{mod}$) in Fig.~\ref{fig:figure1} (b). This can be understood by considering the frequency dependence of the envelope of the peaks, which is determined by the $\sin(\pi f \Delta t)/f$ term of (\ref{eq:2}). For $f \Delta t\ll1$, this term can be written as $\pi \Delta t$, which is independent of the frequency $f$ and therefore all harmonics contribute equally. 

In pEDMR, the photocurrent response typically occurs as a transient which decays within tens of microseconds after the mw pulses~\cite{Boehme03EDMR}, while typical shot repetition times are 1~ms and therefore $f_\mathrm{mod} \Delta t=\Delta t/(2\tau_\mathrm{srt})\approx1/100\ll1$. Therefore, the modulated signal contains frequency components at odd multiples of $f_\mathrm{mod}$ up to a frequency $f\approx 1/\Delta t\approx50$~kHz. For a cut-off frequency of a high-pass filter $f_\mathrm{3dB}=2$~kHz typically used to surpress low-frequency current noise which is larger than the modulation frequency $f_\mathrm{mod}<500$~Hz, the first harmonics are surpressed, but most of the signal at higher harmonics will pass through the filter.     
The width of the peak at the fundamental frequency (as well as for all harmonics) and therefore the bandwidth of the lock-in detection scheme $\Delta f\propto 1/N_\mathrm{cycle}$ and thus $\Delta f\propto 1/T_\mathrm{meas}$, as shown in Fig.~\ref{fig:figure1}(c) for $\Delta t$=1/(10$f_\mathrm{mod}$). 

For an experimental demonstration of this detection scheme, we use Si:P epilayers consisting of a $22$~nm thick Si layer with
a nominal P concentration of $9\times10^{16}$\,cm$^{-3}$, covered with a native
oxide and grown on a $2.5$~\textmu{}m thick, nominally undoped Si buffer on a silicon-on-insulator substrate. EDMR signals observed in this type of sample originate dominantly from spin-dependent recombination between $^{31}$P donors and Si/SiO$_2$ interface states (P$_\mathrm{b0}$)~\cite{Hoehne09KSM}.
For electrical measurements, interdigit Cr/Au contacts with a spacing of $20$~$\mu$m
covering an active area of $2\times2.25$~mm$^{2}$ are evaporated.
All experiments are performed at $\sim$5~K in a dielectric microwave resonator for pulsed EPR at X-band frequencies. The samples are illuminated with above-bandgap light and biased with 100~mV resulting in a current of $\sim$60~$\mu$A. The current transients after the pulse sequence are amplified by a custom-built balanced transimpedance amplifier (Elektronik-Manufaktur, Mahlsdorf) with low- and high-pass filtering at cut-off frequencies of 1~MHz and 2~kHz, respectively.                 
In all experiments, we choose the microwave frequency and external magnetic field such that the microwave pulses resonantly excite the spectrally isolated high-field P hyperfine line~\cite{Hoehne09KSM}. We apply the spin echo pulse sequence with 30~ns long $\pi$ pulses, $N_\mathrm{cycle}$=1000 and a shot repetition time $\tau_\mathrm{srt}$=5~ms resulting in a modulation frequency of $f_\mathrm{mod}$=100~Hz. 

\begin{figure}[ht]
\begin{centering}
\includegraphics[width=8.0cm]{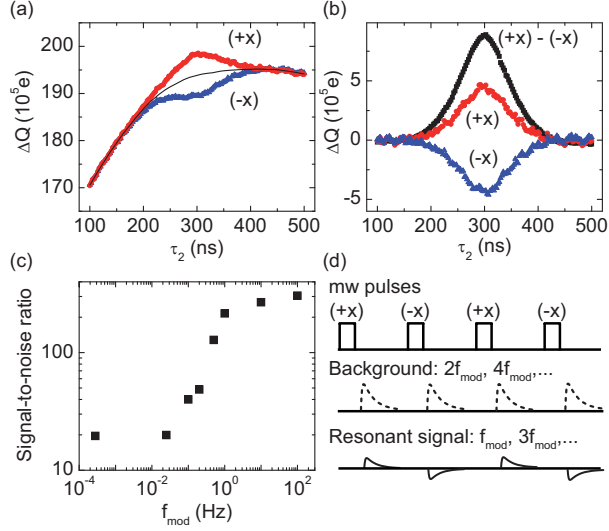}
\par\end{centering}
\caption{\label{fig:figure2}
(a) Integrated charge $\Delta Q$ as a function of $\tau_2$ for $\tau_1$=300~ns measured with phase modulation at $f_\mathrm{mod}=100$~Hz. The data points with the phase of the last $\pi$/2 pulse set to (+x) (upper trace) and (\hbox{-}x) (lower trace) are shown separately. (b) Echo trace obtained by subtracting the two echo traces (+x) and (\hbox{-}x). For comparison, the echo traces (+x) and (\hbox{-}x) after subtraction of the background taken as the smoothed average of the two traces in (a) are shown as well. (c) Signal-to-noise ratio of an electrically detected spin echo as a function of the modulation frequency $f_\mathrm{mod}$. (d) Sketch of the non-resonant (dashed lines) and resonant current transients (solid lines) with Fourier components at even multiples and odd mutiples of $f_\mathrm{mod}$, respectively.}
\end{figure}

In Fig.~\ref{fig:figure2}(a), the integrated charge is shown separately for (+x) and (\hbox{-}x) as a function of $\tau_2$ for $\tau_1$=300~ns. The echo peaks are visible at $\tau_2$=300~ns on top of a large background with positive echo amplitude for (+x) and negative echo amplitude for (\hbox{-}x) while the background is the same for the two phases. To recover the signal, we subtract the two traces from each other resulting in the trace (+x)-(\hbox{-}x) shown in Fig.~\ref{fig:figure2}(b). For comparison, the echo traces (+x) and (\hbox{-}x) after subtraction of the background taken as the smoothed average of the two traces (black line in Fig.\,\ref{fig:figure2}(a)) are shown as well. In addition to the effective removal of the background, comparison of the noise level in traces (+x) and (\hbox{-}x) with their difference (+x)-(\hbox{-}x) illustrates the considerable reduction of noise by the lock-in detection scheme.  

The benefit of this modulation scheme is further demonstrated by measuring the noise as a function of the modulation frequency $f_\mathrm{mod}$. To change the modulation frequency $f_\mathrm{mod}$ independently of the measurement time, in every cycle we repeat the pulse sequence (+x) $N_\mathrm{avr}$-times followed by $N_\mathrm{avr}$ pulse sequences (\hbox{-}x), so that $f_\mathrm{mod}=1/(2N_\mathrm{avr}\tau_\mathrm{srt})$. Varying $N_\mathrm{avr}$ and $N_\mathrm{cycle}$ between 1 and 1000, while keeping the number of sample points $N_\mathrm{avr}\cdot N_\mathrm{cycle}$ constant, changes $f_\mathrm{mod}$ from 0.1~Hz to 100~Hz at a constant bandwidth of $\approx1/T_\mathrm{meas}=$0.1~Hz. The noise is quantified as the standard deviation of 90 measurements of the echo amplitude $\Delta Q$ for $\tau_1$=$\tau_2$=300~ns, where for each measurement $N_\mathrm{avr}\cdot N_\mathrm{cycle}=$1000 sample points are recorded. 

In Fig.~\ref{fig:figure2} (c), the signal-to-noise ratio, obtained by dividing the echo peak amplitude by the standard deviation of the noise defined above, is plotted as a function of $f_\mathrm{mod}$. By increasing the modulation frequency from several mHz to 100~Hz the signal-to-noise ratio is improved by more than one order of magnitude. The data point at $f_\mathrm{mod}$=0.025~Hz is measured with $\tau_\mathrm{srt}=$20~ms, $N_\mathrm{avr}=1000$ and $N_\mathrm{cycle}=1$ resulting in a 4 times longer measurement time $T_\mathrm{meas}$ when compared to the other data points. Since the bandwidth of the lock-in detection scheme $\Delta f\propto1/T_\mathrm{meas}$, the obtained noise amplitude is divided by 2 to make it comparable with the other values. The data point at $f_\mathrm{mod}$=0.3~mHz is taken without phase modulation. In this case, $f_\mathrm{mod}$ is calculated as the inverse of the overall measurement time. 

The signal-to-noise ratio saturates both at low as well as at high frequencies. A measurement of the current noise of the illuminated Si sample without the application of microwave pulses shows that the noise floor at high $f_\mathrm{mod}$ is determined by the current noise of the sample (data not shown). The current noise spectrum shows a strong increase of the noise at low frequencies, which is, however, removed effectively by the high-pass filter of the amplifier, so that a flat noise spectrum is observed at its output. We therefore conclude that the strong decrease of the signal-to-noise ratio at lower frequencies observed in Fig.~\ref{fig:figure2} (c) is due to low-frequency noise of the background current transients induced by the strong microwave pulses. We tentatively attribute this noise to low-frequency variations of the mw pulse amplitude, which, due to the non-resonant current transients, has stronger effects on EDMR measurements when compared to ESR. This noise, although at low-frequencies, is not removed by the high-pass filter as discussed below.
 
Since the amplitude of the non-resonant current transients is independent of the phase of the mw pulse, the background signal contains Fourier components at even multiples of $f_\mathrm{mod}$, while the Fourier components of the signal occur at odd multiples of $f_\mathrm{mod}$ as sketched in Fig.~\ref{fig:figure2}(d). Both signals occur on the same timescale and therefore contain Fourier components up to $\approx 50$~kHz as discussed above.
Noise in the amplitude of the mw pulses at frequencies $f_\mathrm{noise}$ will be mixed with the background signal resulting in noise components at $2f_\mathrm{mod}\pm f_\mathrm{noise}$ and higher even harmonics, which are not filtered out by the high-pass filter. However, the lock-in detection scheme is only sensitive to signals at odd harmonics of $f_\mathrm{mod}$ (see Fig.~\ref{fig:figure1}) and, therefore, the low-frequency noise is removed for large $f_\mathrm{mod}$ as shown in Fig.~\ref{fig:figure2}(c). Since noise at $f_\mathrm{noise}=f_\mathrm{mod}$ cannot be removed by lock-in detection, the signal-to-noise ratio decreases for smaller $f_\mathrm{mod}$ due to the low-frequency noise.        
   
In most pulsed EDMR experiments until now, the large microwave-induced background is removed by measuring additional traces at different values of the static magnetic field where no resonant processes are observed~\cite{Stegner06}. In the approach presented here, no additional traces at off-resonance fields have to be measured since the background is subtracted by the lock-in detection scheme. Since for a spin echo without lock-in detection conventional pEDMR measurements were performed at typically two additional values of the magnetic field, the phase-cycling itself reduces the measurement time by a factor of 3. Together with the tenfold increase of the signal-to-noise ratio due to the lock-in detection, this leads to a reduction of the measurement time by a factor of $\sim$300. In principle, for pulse sequences where phase modulation is not feasible, other parameters like the microwave frequency or the magnetic field can be modulated. 

In summary, we have demonstrated theoretically and experimentally a lock-in detection scheme for pulsed EDMR experiments which significantly improves the signal-to-noise ratio. This scheme allows
to extend the experimental methods of pEDMR to more advanced pulse sequences~\cite{HoehneENDOR2011, HoehneESEEM2011} and opens its application to other materials and spin-dependent processes to be studied with this technique.

This work was supported by DFG
(Grant No.~SFB 631, C3) and BMBF (EPR-Solar, Grant No.~03SF0328).

\bibliographystyle{aipnum4-1}

\begin{thebibliography}{21}%
\makeatletter
\providecommand \@ifxundefined [1]{%
 \@ifx{#1\undefined}
}%
\providecommand \@ifnum [1]{%
 \ifnum #1\expandafter \@firstoftwo
 \else \expandafter \@secondoftwo
 \fi
}%
\providecommand \@ifx [1]{%
 \ifx #1\expandafter \@firstoftwo
 \else \expandafter \@secondoftwo
 \fi
}%
\providecommand \natexlab [1]{#1}%
\providecommand \enquote  [1]{``#1''}%
\providecommand \bibnamefont  [1]{#1}%
\providecommand \bibfnamefont [1]{#1}%
\providecommand \citenamefont [1]{#1}%
\providecommand \href@noop [0]{\@secondoftwo}%
\providecommand \href [0]{\begingroup \@sanitize@url \@href}%
\providecommand \@href[1]{\@@startlink{#1}\@@href}%
\providecommand \@@href[1]{\endgroup#1\@@endlink}%
\providecommand \@sanitize@url [0]{\catcode `\\12\catcode `\$12\catcode
  `\&12\catcode `\#12\catcode `\^12\catcode `\_12\catcode `\%12\relax}%
\providecommand \@@startlink[1]{}%
\providecommand \@@endlink[0]{}%
\providecommand \url  [0]{\begingroup\@sanitize@url \@url }%
\providecommand \@url [1]{\endgroup\@href {#1}{\urlprefix }}%
\providecommand \urlprefix  [0]{URL }%
\providecommand \Eprint [0]{\href }%
\providecommand \doibase [0]{http://dx.doi.org/}%
\providecommand \selectlanguage [0]{\@gobble}%
\providecommand \bibinfo  [0]{\@secondoftwo}%
\providecommand \bibfield  [0]{\@secondoftwo}%
\providecommand \translation [1]{[#1]}%
\providecommand \BibitemOpen [0]{}%
\providecommand \bibitemStop [0]{}%
\providecommand \bibitemNoStop [0]{.\EOS\space}%
\providecommand \EOS [0]{\spacefactor3000\relax}%
\providecommand \BibitemShut  [1]{\csname bibitem#1\endcsname}%
\let\auto@bib@innerbib\@empty
\bibitem [{\citenamefont {Spaeth}\ and\ \citenamefont
  {Overhof}(2003)}]{Spaeth03}%
  \BibitemOpen
  \bibfield  {author} {\bibinfo {author} {\bibfnamefont {J.-M.}\ \bibnamefont
  {Spaeth}}\ and\ \bibinfo {author} {\bibfnamefont {H.}~\bibnamefont
  {Overhof}},\ }\href@noop {} {\emph {\bibinfo {title} {Point Defects in
  Semiconductors and Insulators}}}\ (\bibinfo  {publisher} {Springer, Berlin},\
  \bibinfo {year} {2003})\BibitemShut {NoStop}%
\bibitem [{\citenamefont {Maier}(1997)}]{Maier97}%
  \BibitemOpen
  \bibfield  {author} {\bibinfo {author} {\bibfnamefont {D.~C.}\ \bibnamefont
  {Maier}},\ }\href@noop {} {\bibfield  {journal} {\bibinfo  {journal} {Bruker
  Rep.}\ }\textbf {\bibinfo {volume} {144}},\ \bibinfo {pages} {13} (\bibinfo
  {year} {1997})}\BibitemShut {NoStop}%
\bibitem [{\citenamefont {Lepine}(1972)}]{Lepine72Spindep}%
  \BibitemOpen
  \bibfield  {author} {\bibinfo {author} {\bibfnamefont {D.~J.}\ \bibnamefont
  {Lepine}},\ }\href {\doibase 10.1103/PhysRevB.6.436} {\bibfield  {journal}
  {\bibinfo  {journal} {Phys. Rev. B}\ }\textbf {\bibinfo {volume} {6}},\
  \bibinfo {pages} {436} (\bibinfo {year} {1972})}\BibitemShut {NoStop}%
\bibitem [{\citenamefont {Frankevich}, \citenamefont {Pristupa},\ and\
  \citenamefont {Kobryanskii}(1984)}]{Frankevich84}%
  \BibitemOpen
  \bibfield  {author} {\bibinfo {author} {\bibfnamefont {E.~L.}\ \bibnamefont
  {Frankevich}}, \bibinfo {author} {\bibfnamefont {A.~I.}\ \bibnamefont
  {Pristupa}}, \ and\ \bibinfo {author} {\bibfnamefont {V.~M.}\ \bibnamefont
  {Kobryanskii}},\ }\href@noop {} {\bibfield  {journal} {\bibinfo  {journal}
  {JETP Lett.}\ }\textbf {\bibinfo {volume} {40}},\ \bibinfo {pages} {733}
  (\bibinfo {year} {1984})}\BibitemShut {NoStop}%
\bibitem [{\citenamefont {Dyakonov}\ \emph {et~al.}(1994)\citenamefont
  {Dyakonov}, \citenamefont {Gauss}, \citenamefont {R\"osler}, \citenamefont
  {Karg}, \citenamefont {Rie\ss},\ and\ \citenamefont
  {Schwoerer}}]{Dyakonov1994}%
  \BibitemOpen
  \bibfield  {author} {\bibinfo {author} {\bibfnamefont {V.}~\bibnamefont
  {Dyakonov}}, \bibinfo {author} {\bibfnamefont {N.}~\bibnamefont {Gauss}},
  \bibinfo {author} {\bibfnamefont {G.}~\bibnamefont {R\"osler}}, \bibinfo
  {author} {\bibfnamefont {S.}~\bibnamefont {Karg}}, \bibinfo {author}
  {\bibfnamefont {W.}~\bibnamefont {Rie\ss}}, \ and\ \bibinfo {author}
  {\bibfnamefont {M.}~\bibnamefont {Schwoerer}},\ }\href {\doibase
  10.1016/0301-0104(94)00340-8} {\bibfield  {journal} {\bibinfo  {journal}
  {Chem. Phys.}\ }\textbf {\bibinfo {volume} {189}},\ \bibinfo {pages}
  {687} (\bibinfo {year} {1994})}\BibitemShut {NoStop}%
\bibitem [{\citenamefont {Carlos}\ and\ \citenamefont
  {Nakamura}(1997)}]{Carlos1997}%
  \BibitemOpen
  \bibfield  {author} {\bibinfo {author} {\bibfnamefont {W.~E.}\ \bibnamefont
  {Carlos}}\ and\ \bibinfo {author} {\bibfnamefont {S.}~\bibnamefont
  {Nakamura}},\ }\href {\doibase 10.1063/1.118778} {\bibfield  {journal}
  {\bibinfo  {journal} {Appl. Phys. Lett.}\ }\textbf {\bibinfo {volume} {70}},\
  \bibinfo {pages} {2019} (\bibinfo {year} {1997})}\BibitemShut {NoStop}%
\bibitem [{\citenamefont {Stutzmann}, \citenamefont {Brandt},\ and\
  \citenamefont {Bayerl}(2000)}]{Stutzmann2000}%
  \BibitemOpen
  \bibfield  {author} {\bibinfo {author} {\bibfnamefont {M.}~\bibnamefont
  {Stutzmann}}, \bibinfo {author} {\bibfnamefont {M.~S.}\ \bibnamefont
  {Brandt}}, \ and\ \bibinfo {author} {\bibfnamefont {M.~W.}\ \bibnamefont
  {Bayerl}},\ }\href {\doibase 10.1016/S0022-3093(99)00871-6} {\bibfield
  {journal} {\bibinfo  {journal} {J. {Non-Cryst.} Solids}\ }\textbf {\bibinfo
  {volume} {266-269}},\ \bibinfo {pages} {1} (\bibinfo {year}
  {2000})}\BibitemShut {NoStop}%
\bibitem [{\citenamefont {Graeff}\ \emph {et~al.}(2005)\citenamefont {Graeff},
  \citenamefont {Silva}, \citenamefont {N\"uesch},\ and\ \citenamefont
  {Zuppiroli}}]{Graeff2005}%
  \BibitemOpen
  \bibfield  {author} {\bibinfo {author} {\bibfnamefont {C.~F.~O.}\
  \bibnamefont {Graeff}}, \bibinfo {author} {\bibfnamefont {G.~B.}\
  \bibnamefont {Silva}}, \bibinfo {author} {\bibfnamefont {F.}~\bibnamefont
  {N\"uesch}}, \ and\ \bibinfo {author} {\bibfnamefont {L.}~\bibnamefont
  {Zuppiroli}},\ }\href {\doibase 10.1140/epje/i2005-10026-6} {\bibfield
  {journal} {\bibinfo  {journal} {Eur. Phys. J. E}\ }\textbf {\bibinfo {volume}
  {18}},\ \bibinfo {pages} {8} (\bibinfo {year} {2005})}\BibitemShut {NoStop}%
\bibitem [{\citenamefont {Schweiger}\ and\ \citenamefont
  {Jeschke}(2001)}]{Schweiger01}%
  \BibitemOpen
  \bibfield  {author} {\bibinfo {author} {\bibfnamefont {A.}~\bibnamefont
  {Schweiger}}\ and\ \bibinfo {author} {\bibfnamefont {G.}~\bibnamefont
  {Jeschke}},\ }\href@noop {} {\emph {\bibinfo {title} {Principles of pulse
  electron paramagnetic resonance}}}\ (\bibinfo  {publisher} {Oxford
  {U}niversity {P}ress, {O}xford},\ \bibinfo {year} {2001})\BibitemShut
  {NoStop}%
\bibitem [{\citenamefont {Boehme}\ and\ \citenamefont
  {Lips}(2003)}]{Boehme03EDMR}%
  \BibitemOpen
  \bibfield  {author} {\bibinfo {author} {\bibfnamefont {C.}~\bibnamefont
  {Boehme}}\ and\ \bibinfo {author} {\bibfnamefont {K.}~\bibnamefont {Lips}},\
  }\href {\doibase 10.1103/PhysRevB.68.245105} {\bibfield  {journal} {\bibinfo
  {journal} {Phys. Rev. B}\ }\textbf {\bibinfo {volume} {68}},\ \bibinfo
  {pages} {245105} (\bibinfo {year} {2003})}\BibitemShut {NoStop}%
\bibitem [{\citenamefont {Stegner}\ \emph {et~al.}(2006)\citenamefont
  {Stegner}, \citenamefont {Boehme}, \citenamefont {Huebl}, \citenamefont
  {Stutzmann}, \citenamefont {Lips},\ and\ \citenamefont {Brandt}}]{Stegner06}%
  \BibitemOpen
  \bibfield  {author} {\bibinfo {author} {\bibfnamefont {A.~R.}\ \bibnamefont
  {Stegner}}, \bibinfo {author} {\bibfnamefont {C.}~\bibnamefont {Boehme}},
  \bibinfo {author} {\bibfnamefont {H.}~\bibnamefont {Huebl}}, \bibinfo
  {author} {\bibfnamefont {M.}~\bibnamefont {Stutzmann}}, \bibinfo {author}
  {\bibfnamefont {K.}~\bibnamefont {Lips}}, \ and\ \bibinfo {author}
  {\bibfnamefont {M.~S.}\ \bibnamefont {Brandt}},\ }\href@noop {} {\bibfield
  {journal} {\bibinfo  {journal} {Nat. Physics}\ }\textbf {\bibinfo {volume}
  {2}},\ \bibinfo {pages} {835} (\bibinfo {year} {2006})}\BibitemShut {NoStop}%
\bibitem [{\citenamefont {Harneit}\ \emph {et~al.}(2007)\citenamefont
  {Harneit}, \citenamefont {Boehme}, \citenamefont {Schaefer}, \citenamefont
  {Huebener}, \citenamefont {Fostiropoulos},\ and\ \citenamefont
  {Lips}}]{Harneit2007}%
  \BibitemOpen
  \bibfield  {author} {\bibinfo {author} {\bibfnamefont {W.}~\bibnamefont
  {Harneit}}, \bibinfo {author} {\bibfnamefont {C.}~\bibnamefont {Boehme}},
  \bibinfo {author} {\bibfnamefont {S.}~\bibnamefont {Schaefer}}, \bibinfo
  {author} {\bibfnamefont {K.}~\bibnamefont {Huebener}}, \bibinfo {author}
  {\bibfnamefont {K.}~\bibnamefont {Fostiropoulos}}, \ and\ \bibinfo {author}
  {\bibfnamefont {K.}~\bibnamefont {Lips}},\ }\href {\doibase
  10.1103/PhysRevLett.98.216601} {\bibfield  {journal} {\bibinfo  {journal}
  {Phys. Rev. Lett.}\ }\textbf {\bibinfo {volume} {98}},\ \bibinfo {pages}
  {216601} (\bibinfo {year} {2007})}\BibitemShut {NoStop}%
\bibitem [{\citenamefont {McCamey}\ \emph {et~al.}(2008)\citenamefont
  {McCamey}, \citenamefont {Seipel}, \citenamefont {Paik}, \citenamefont
  {Walter}, \citenamefont {Borys}, \citenamefont {Lupton},\ and\ \citenamefont
  {Boehme}}]{McCamey08}%
  \BibitemOpen
  \bibfield  {author} {\bibinfo {author} {\bibfnamefont {D.~R.}\ \bibnamefont
  {McCamey}}, \bibinfo {author} {\bibfnamefont {H.~A.}\ \bibnamefont {Seipel}},
  \bibinfo {author} {\bibfnamefont {S.-Y.}\ \bibnamefont {Paik}}, \bibinfo
  {author} {\bibfnamefont {M.~J.}\ \bibnamefont {Walter}}, \bibinfo {author}
  {\bibfnamefont {N.~J.}\ \bibnamefont {Borys}}, \bibinfo {author}
  {\bibfnamefont {J.~M.}\ \bibnamefont {Lupton}}, \ and\ \bibinfo {author}
  {\bibfnamefont {C.}~\bibnamefont {Boehme}},\ }\href@noop {} {\bibfield
  {journal} {\bibinfo  {journal} {Nat. Mat.}\ }\textbf {\bibinfo {volume}
  {7}},\ \bibinfo {pages} {723} (\bibinfo {year} {2008})}\BibitemShut {NoStop}%
\bibitem [{\citenamefont {Behrends}\ \emph {et~al.}(2009)\citenamefont
  {Behrends}, \citenamefont {Schnegg}, \citenamefont {Fehr}, \citenamefont
  {Lambertz}, \citenamefont {Haas}, \citenamefont {Finger}, \citenamefont
  {Rech},\ and\ \citenamefont {Lips}}]{Behrends09II}%
  \BibitemOpen
  \bibfield  {author} {\bibinfo {author} {\bibfnamefont {J.}~\bibnamefont
  {Behrends}}, \bibinfo {author} {\bibfnamefont {A.}~\bibnamefont {Schnegg}},
  \bibinfo {author} {\bibfnamefont {M.}~\bibnamefont {Fehr}}, \bibinfo {author}
  {\bibfnamefont {A.}~\bibnamefont {Lambertz}}, \bibinfo {author}
  {\bibfnamefont {S.}~\bibnamefont {Haas}}, \bibinfo {author} {\bibfnamefont
  {F.}~\bibnamefont {Finger}}, \bibinfo {author} {\bibfnamefont
  {B.}~\bibnamefont {Rech}}, \ and\ \bibinfo {author} {\bibfnamefont
  {K.}~\bibnamefont {Lips}},\ }\href@noop {} {\bibfield  {journal} {\bibinfo
  {journal} {Phil. Mag.}\ }\textbf {\bibinfo {volume} {89}},\ \bibinfo {pages}
  {2655} (\bibinfo {year} {2009})}\BibitemShut {NoStop}%
\bibitem [{\citenamefont {Behrends}\ \emph {et~al.}(2010)\citenamefont
  {Behrends}, \citenamefont {Schnegg}, \citenamefont {Lips}, \citenamefont
  {Thomsen}, \citenamefont {Pandey}, \citenamefont {Samuel},\ and\
  \citenamefont {Keeble}}]{Behrends10}%
  \BibitemOpen
  \bibfield  {author} {\bibinfo {author} {\bibfnamefont {J.}~\bibnamefont
  {Behrends}}, \bibinfo {author} {\bibfnamefont {A.}~\bibnamefont {Schnegg}},
  \bibinfo {author} {\bibfnamefont {K.}~\bibnamefont {Lips}}, \bibinfo {author}
  {\bibfnamefont {E.~A.}\ \bibnamefont {Thomsen}}, \bibinfo {author}
  {\bibfnamefont {A.~K.}\ \bibnamefont {Pandey}}, \bibinfo {author}
  {\bibfnamefont {I.~D.~W.}\ \bibnamefont {Samuel}}, \ and\ \bibinfo {author}
  {\bibfnamefont {D.~J.}\ \bibnamefont {Keeble}},\ }\href {\doibase
  10.1103/PhysRevLett.105.176601} {\bibfield  {journal} {\bibinfo  {journal}
  {Phys. Rev. Lett.}\ }\textbf {\bibinfo {volume} {105}},\ \bibinfo {pages}
  {176601} (\bibinfo {year} {2010})}\BibitemShut {NoStop}%
\bibitem [{\citenamefont {Hoehne}\ \emph
  {et~al.}(2011{\natexlab{a}})\citenamefont {Hoehne}, \citenamefont {Dreher},
  \citenamefont {Huebl}, \citenamefont {Stutzmann},\ and\ \citenamefont
  {Brandt}}]{HoehneENDOR2011}%
  \BibitemOpen
  \bibfield  {author} {\bibinfo {author} {\bibfnamefont {F.}~\bibnamefont
  {Hoehne}}, \bibinfo {author} {\bibfnamefont {L.}~\bibnamefont {Dreher}},
  \bibinfo {author} {\bibfnamefont {H.}~\bibnamefont {Huebl}}, \bibinfo
  {author} {\bibfnamefont {M.}~\bibnamefont {Stutzmann}}, \ and\ \bibinfo
  {author} {\bibfnamefont {M.~S.}\ \bibnamefont {Brandt}},\ }\href {\doibase
  10.1103/PhysRevLett.106.187601} {\bibfield  {journal} {\bibinfo  {journal}
  {Phys. Rev. Lett.}\ }\textbf {\bibinfo {volume} {106}},\ \bibinfo {pages}
  {187601} (\bibinfo {year} {2011}{\natexlab{a}})}\BibitemShut {NoStop}%
\bibitem [{\citenamefont {Hoehne}\ \emph
  {et~al.}(2011{\natexlab{b}})\citenamefont {Hoehne}, \citenamefont {Lu},
  \citenamefont {Stegner}, \citenamefont {Stutzmann}, \citenamefont {Brandt},
  \citenamefont {Rohrm\"uller}, \citenamefont {Schmidt},\ and\ \citenamefont
  {Gerstmann}}]{HoehneESEEM2011}%
  \BibitemOpen
  \bibfield  {author} {\bibinfo {author} {\bibfnamefont {F.}~\bibnamefont
  {Hoehne}}, \bibinfo {author} {\bibfnamefont {J.}~\bibnamefont {Lu}}, \bibinfo
  {author} {\bibfnamefont {A.~R.}\ \bibnamefont {Stegner}}, \bibinfo {author}
  {\bibfnamefont {M.}~\bibnamefont {Stutzmann}}, \bibinfo {author}
  {\bibfnamefont {M.~S.}\ \bibnamefont {Brandt}}, \bibinfo {author}
  {\bibfnamefont {M.}~\bibnamefont {Rohrm\"uller}}, \bibinfo {author}
  {\bibfnamefont {W.~G.}\ \bibnamefont {Schmidt}}, \ and\ \bibinfo {author}
  {\bibfnamefont {U.}~\bibnamefont {Gerstmann}},\ }\href {\doibase
  10.1103/PhysRevLett.106.196101} {\bibfield  {journal} {\bibinfo  {journal}
  {Phys. Rev. Lett.}\ }\textbf {\bibinfo {volume} {106}},\ \bibinfo {pages}
  {196101} (\bibinfo {year} {2011}{\natexlab{b}})}\BibitemShut {NoStop}%
 \bibitem [{\citenamefont {Fehr}\ \emph {et~al.}(2011)\citenamefont {Fehr},
  \citenamefont {Behrends}, \citenamefont {Haas}, \citenamefont {Rech},
  \citenamefont {Lips},\ and\ \citenamefont {Schnegg}}]{Fehr2011}%
  \BibitemOpen
  \bibfield  {author} {\bibinfo {author} {\bibfnamefont {M.}~\bibnamefont
  {Fehr}}, \bibinfo {author} {\bibfnamefont {J.}~\bibnamefont {Behrends}},
  \bibinfo {author} {\bibfnamefont {S.}~\bibnamefont {Haas}}, \bibinfo {author}
  {\bibfnamefont {B.}~\bibnamefont {Rech}}, \bibinfo {author} {\bibfnamefont
  {K.}~\bibnamefont {Lips}}, \ and\ \bibinfo {author} {\bibfnamefont
  {A.}~\bibnamefont {Schnegg}},\ }\href {\doibase 10.1103/PhysRevB.84.193202}
  {\bibfield  {journal} {\bibinfo  {journal} {Phys. Rev. B}\ }\textbf {\bibinfo
  {volume} {84}},\ \bibinfo {pages} {193202} (\bibinfo {year}
  {2011})}\BibitemShut {NoStop}%
\bibitem [{\citenamefont {Percival}\ and\ \citenamefont
  {Hyde}(1975)}]{Percival75}%
  \BibitemOpen
  \bibfield  {author} {\bibinfo {author} {\bibfnamefont {P.}~\bibnamefont
  {Percival}}\ and\ \bibinfo {author} {\bibfnamefont {J.~S.}\ \bibnamefont
  {Hyde}},\ }\href@noop {} {\bibfield  {journal} {\bibinfo  {journal} {Rev.
  Sci. Instr.}\ }\textbf {\bibinfo {volume} {46}},\ \bibinfo {pages} {1522}
  (\bibinfo {year} {1975})}\BibitemShut {NoStop}%
\bibitem [{\citenamefont {Dicke}(1946)}]{Dicke1946}%
  \BibitemOpen
  \bibfield  {author} {\bibinfo {author} {\bibfnamefont {R.~H.}\ \bibnamefont
  {Dicke}},\ }\href {\doibase 10.1063/1.1770483} {\bibfield  {journal}
  {\bibinfo  {journal} {Rev. Sci. Instr.}\ }\textbf {\bibinfo {volume} {17}},\
  \bibinfo {pages} {268} (\bibinfo {year} {1946})}\BibitemShut {NoStop}%
\bibitem [{\citenamefont {Huebl}\ \emph {et~al.}(2008)\citenamefont {Huebl},
  \citenamefont {Hoehne}, \citenamefont {Grolik}, \citenamefont {Stegner},
  \citenamefont {Stutzmann},\ and\ \citenamefont {Brandt}}]{Huebl08Echo}%
  \BibitemOpen
  \bibfield  {author} {\bibinfo {author} {\bibfnamefont {H.}~\bibnamefont
  {Huebl}}, \bibinfo {author} {\bibfnamefont {F.}~\bibnamefont {Hoehne}},
  \bibinfo {author} {\bibfnamefont {B.}~\bibnamefont {Grolik}}, \bibinfo
  {author} {\bibfnamefont {A.~R.}\ \bibnamefont {Stegner}}, \bibinfo {author}
  {\bibfnamefont {M.}~\bibnamefont {Stutzmann}}, \ and\ \bibinfo {author}
  {\bibfnamefont {M.~S.}\ \bibnamefont {Brandt}},\ }\href {\doibase
  10.1103/PhysRevLett.100.177602} {\bibfield  {journal} {\bibinfo  {journal}
  {Phys. Rev. Lett.}\ }\textbf {\bibinfo {volume} {100}},\ \bibinfo {eid}
  {177602} (\bibinfo {year} {2008})}\BibitemShut {NoStop}%
\bibitem [{\citenamefont {Hoehne}\ \emph {et~al.}(2010)\citenamefont {Hoehne},
  \citenamefont {Huebl}, \citenamefont {Galler}, \citenamefont {Stutzmann},\
  and\ \citenamefont {Brandt}}]{Hoehne09KSM}%
  \BibitemOpen
  \bibfield  {author} {\bibinfo {author} {\bibfnamefont {F.}~\bibnamefont
  {Hoehne}}, \bibinfo {author} {\bibfnamefont {H.}~\bibnamefont {Huebl}},
  \bibinfo {author} {\bibfnamefont {B.}~\bibnamefont {Galler}}, \bibinfo
  {author} {\bibfnamefont {M.}~\bibnamefont {Stutzmann}}, \ and\ \bibinfo
  {author} {\bibfnamefont {M.~S.}\ \bibnamefont {Brandt}},\ }\href {\doibase
  10.1103/PhysRevLett.104.046402} {\bibfield  {journal} {\bibinfo  {journal}
  {Phys. Rev. Lett.}\ }\textbf {\bibinfo {volume} {104}},\ \bibinfo {pages}
  {046402} (\bibinfo {year} {2010})}\BibitemShut {NoStop}%
\end{thebibliography}

\end{document}